# Multi-Roller Structure Triboelectric Nanogenerator for Enhanced Water Wave Energy Harvesting and Energy Management

Kequan Xia[1, 2], Zhiwei Xu[2*], Lizhong Wang[2*], Min Yu[3*]

[1]Department of Materials Science and Engineering, National University of Singapore, Singapore 117575, Singapore;
[2]Ocean College, Zhejiang University, Zhoushan, 316021, China;
[3]Department of Mechanical Engineering, Imperial College London, SW7 2AZ, London, United Kingdom;

*Corresponding author: xuzw@zju.edu.cn, wanglz@zju.edu.cn, m.yu14@imperial.ac.uk.

*Abstract*—Wave energy harvesting is critical for advancing the development and utilization of marine resources. In this study, we present a novel multi-roller structure triboelectric nanogenerator (MR-TENG) designed specifically for efficient water wave energy harvesting. The MR-TENG leverages a coupled multi-roller design to significantly enhance its energy harvesting capabilities. The triboelectric layers are composed of polytetrafluoroethylene (PTFE) film and paper, with a grid copper electrode serving as the conductive element. Through an optimized energy output strategy, a single MR-TENG is capable of generating 602.045 μJ of electrical energy within 100 s. The device achieves a short-circuit current ($I_{sc}$) of approximately 2.06 μA and an open-circuit voltage ($V_{oc}$) of around 166 V. We further investigate the impact of different connection modes, including parallel and series configurations, on the performance of MR-TENG arrays. Notably, the electrical energy produced by the MR-TENG array is sufficient to power 40 blue commercial light-emitting diodes (LEDs). This research not only introduces a versatile optimization approach and energy management strategy for roller-structured TENGs but also contributes significantly to the advancement of ocean-based TENG technology.

## 1. Introduction

With the increasing consumption of fossil fuels by industry, the energy crisis and environmental pollution issues are becoming increasingly prominent. Thus, exploring sustainable renewable energy is an effective way to alleviate these challenges. At present, advanced power generation technologies for obtaining renewable energy (solar, wind, and geothermal) are still affected and limited by the environment and equipment. The ocean is vast and rich in resources, and marine resource development technology is closely related to human society's progress [1]. The conventional use of fossil fuels can bring environmental pollution, and meanwhile, limited fossil fuels are challenging to meet the needs of human social development [2, 3]. The marine environment is rich in mechanical energy (such as wave power, tidal power, wind energy, etc.), which has the advantages of wide distribution, large reserves, green, etc., and will become an essential substitute for fossil energy in the future [4-7]. As one of the most promising energy sources in the natural environment, water wave energy provides an opportunity to alleviate the pressure of depletion of fossil energy reserves due to its wide distribution, large quantity, and clean and renewable characteristics. Currently, most power generation equipment in the ocean still relies on electromagnetic generators (EMGs) technology [8], which has problems such as high construction and maintenance costs, low-frequency energy conversion efficiency, and poor reliability and stability of wave energy collection, limiting their large-scale application and development. Usually, additional mechanisms are needed to convert low-frequency mechanical energy into high-frequency mechanical energy to achieve power generation effects, which limits the harvesting of low-frequency, random, and low-amplitude ocean wave energy. Additionally, EMG equipment is bulky and heavy, which can incur high costs during long-term maintenance. Besides, due to the demand for the Marine Internet of Things (M-IoTs), distributed sensor network nodes used for ocean situational awareness are facing continuous power supply problems, which require a new era of power generation technology suitable for the marine environment [9]. In addition to EMG, moisture nanogenerator [10], piezoelectric nanogenerator (PENG) [11], triboelectric nanogenerator (TENG) [12], and thermoelectric nanogenerator [13] have been explored for energy harvesting, sensor monitoring, and other fields. The working mechanism of the TENG device is based on triboelectrification and electrostatic induction [14-16]. The TENG devices can capture various low-frequency, disordered micro-mechanical energy [17-20]. It can also sense mechanical motion characteristics [21]. The raw materials used for preparing TENG devices are diverse and cost-effective, and the rich TENG structures are widely used in blue energy [22, 23], self-powered sensing [24], and high-voltage electrical energy [25]. EMG equipment is suitable for the conversion of high-frequency and regular kinetic energy. Due to the low-frequency and large-scale distributed characteristics of ocean waves, as well as the randomness of ocean wave characteristics, it is difficult for EMG to efficiently capture and harvest wave energy. In addition, most EMG acquisition devices require multiple levels of energy conversion, and wave energy undergoes varying degrees of loss during each level of conversion. Furthermore, the layout and construction of the electromagnetic induction generator acquisition device is relatively difficult, and daily maintenance costs are high. Compared to traditional EMG power generation equipment, TENG devices have outstanding advantages in terms of structure and materials. In particular, TENG devices are light in weight and compact in size, and are deemed as a significant technology for low-frequency ocean energy harvesting.

Currently, TENG has made progress in ocean blue energy harvesting, mainly including material modification, structural optimization, energy management, and other aspects. For example, Cheng et al. proposed a material and structural improvement approach to significantly increase the spherical TENGs output performance [26]. Liang et al. built a hexagonal TENG network through the integrated energy management module to enhance the power generation efficiency of water wave energy TENG [27]. Lei et al. designed a butterfly-inspired TENG to obtain multi-directional water wave energy [28]. Furthermore, Wu et al. reported a hybridized triboelectric-electromagnetic nanogenerator based on a magnetic sphere for water wave energy harvesting from any direction [29]. Zhang et al. proposed a self-powered TENG water level sensor on ocean vessels according to the liquid-solid triboelectrification mechanism [30]. Feng et al. proposed a soft-contact cylindrical hybridized triboelectric-electromagnetic nanogenerator for ultra-low frequency wave energy harvesting [31]. Cao et al. reported a TENG enabled by coupling the swing-rotation switching mechanism achieves broadband and output-controllable [32]. Han et al. fabricated a cylindrical wave-driven linkage mechanism TENG for effectively harvesting water wave energy [33]. Recently, Hu et al. proposed a wheel-structured TENG with a hyperelastic network, which can efficiently harvest small amplitude wave energy [34]. Compared to traditional EMG equipment, TENG devices have outstanding advantages in terms of structure and materials. At present, relevant research mainly focuses on the design of TENG structures and the improvement of triboelectric material properties, with little consideration given to the coupling issues between composite structures, including interference between output units, device space utilization, and friction efficiency between triboelectric layers. Obviously, establishing a coupling mechanism between the various work units of TENG and achieving high energy output density is crucial for the development of TENG. Thus, the roller structure ocean TENG device can efficiently convert wave power into electric energy, demonstrating excellent application potential. However, the roller structure TENG has a low space utilization rate, and thus, there is an urgent need for a compact structural design strategy to improve TENG performance.

Herein, we report a multi-roller structure triboelectric nanogenerator (MR-TENG) for water wave energy harvesting. Compared to the previous roller structure TENG, the MR-TENG achieved higher space utilization and brought a new path for the TENG design of the roller structure. The MR-TENG's triboelectric layers are composed of polytetrafluoroethylene (PTFE) film and paper. Meanwhile, the design of grid electrodes can significantly improve the probability of friction between rollers to enhance mechanical energy harvesting efficiency. Besides, the connection method between MR-TENG work units will determine the output energy of MR-TENG, such as rectification followed by parallel connection (RPC) mode and parallel connection followed by rectification (PCR) mode. The short-circuit current ($I_{sc}$) and open-circuit voltage ($V_{oc}$) of MR-TENG can arrive at ~2.06 μA and ~166 V, respectively. The connection mode (parallel and series) of the MR-TENG array can bring different output effects. In parallel mode, the output energy efficiency of the MR-TENG array is relatively high. In series mode, the $V_{oc}$ of the MR-TENG array is higher. The MR-TENG array can power 40 blue commercial light-emitting diodes (LEDs) under micro-vibration.

## 2. Experiments

As shown in Fig. 1(a), the roller structure TENG has high potential application value in ocean wave harvesting, and corresponding structure design optimization can help enhance TENG performance. Due to the low-frequency characteristics of waves, the roller structure can effectively capture wave motion and still capture mechanical energy in relatively low amplitude mechanical vibration environments. However, due to the fact that the friction between the rollers belongs to line contact friction, which limits the generation of triboelectric charges, increasing the number of rollers to improve space utilization and achieve higher efficiency of triboelectric charge generation will help to harvest wave energy. Meanwhile, due to the variability of friction between rollers, the introduction of grid electrodes can effectively convert the mechanical energy of the roller itself into frictional electrical energy for a longer time. Accordingly, we have fabricated an MR-TENG integrated with multiple rollers of different sizes, and the triboelectric layer inside the roller is equipped with grid network electrodes to improve the triboelectric efficiency, as exhibited in Fig. 1(b). The PTFE film and paper form the triboelectric functional layers, and the grid copper layer (conductive copper paste) brushed on the paper surface is used for conductive electrodes. The MR-TENG consists of 4 rollers with a length of 10 cm and a thickness of 2 mm. The inner diameters of 4 rollers are 4.5 cm, 3 cm, 2 cm, and 1 cm, respectively. The width of the grid electrode is 5 mm, and the triboelectric layers (PTFE film@ paper) are alternately arranged on the inner and outer walls of the rollers. The high-density grid electrode provides an opportunity for efficient triboelectric contact. Fig. 1(c) displays the picture of MR-TENG with 4 rollers, and three working units can provide higher energy output. There is a gap between the 4 rollers, which can generate contact separation motion when the device is subjected to vibration, thereby generating electrical energy. The inset in Fig. 1(c) illustrates the scanning electron microscope (SEM) image of PTFE film surface. Additionally, the MR-TENG device was installed on the linear motor sports end, and the linear motor system can serve as a power source to provide power for MR-TENG with adjustable frequency and amplitude of movement. The digital oscilloscope (TBS-2000) was connected directly to the MR-TENG output terminal to measure its output under rectified and non-rectified conditions. Keithley 6517B was connected directly to the MR-TENG output terminal to measure the output current and rectified output current of MR-TENG. Moreover, the Keithley 6517B was connected to the both ends of the commercial capacitors to measure the voltage on the commercial capacitors charged by MR-TENG with different forms of rectification circuits.

## 3. Results and Discussion

*3.1. The working mechanism of MR-TENG.*

It is worth noting that during the working process of MR-TENG, there are two friction modes between the triboelectric layers: sliding and rolling, which will be beneficial for the accumulation of triboelectric charges in MR-TENG. Moreover, due to the presence of multiple triboelectric layers in MR-TENG during the working process, to facilitate the description of its working mechanism, we simplified MR-TENG to a double-rolling structure, as illustrated in Fig. 1(d). Additionally, to simplify the description of charge transfer between roller electrodes, we will mainly discuss the triboelectric charges in the triboelectric layer region corresponding to the conductive grid electrodes. After multiple frictions, the distribution of triboelectric charge in two surfaces tends to stabilize, starting from the situation shown in Fig. 1(d1). Driven by external forces in Fig. 2(d2), the internal roller will roll, causing separation between the triboelectric layers with the grid electrode to generate induced current in the external circuit. As the internal roller continues to roll, the triboelectric layer area with grid electrode contacts again to produce a reverse-induced current in the circuit (Fig. 1(d3)). There is not only rolling friction but also sliding friction between the two rollers, which also drives the area corresponding to the grid electrode to have higher frequency contact and separation. As present in Fig. 1(d4), when the corresponding area of the grid electrode is separated due to relative sliding, induced current will be produced again in the circuit. During the continuous generation of rolling and sliding, MR-TENG can generate a continuous current. Meanwhile, owning to the high-density grid electrode, the triboelectric layer will generate multiple current signals during the triboelectric process.

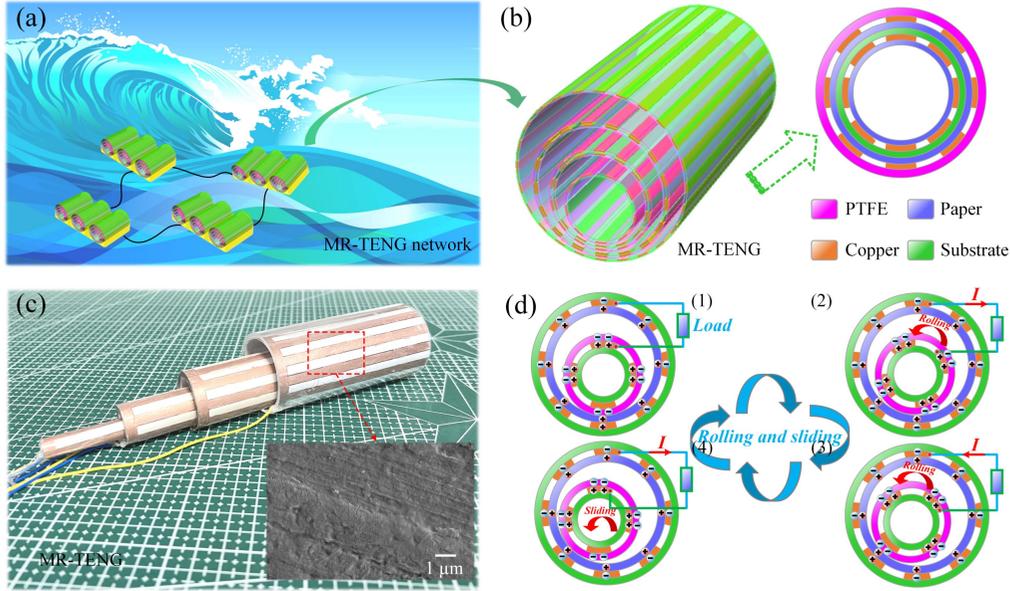

Fig. 1. (a) Illustration of the MR-TENG harvesting wave energy in marine environment. (b) Structural layout diagram of MR-TENG. (c) Physical picture of MR-TENG with four rollers, inset: the SEM image of PTFE surface. (d1-d4) The MR-TENG working principle.

*3.2. The output mode optimization of MR-TENG.*

The MR-TENG with four rollers can form three working units numbered No. 1, No. 2, and No. 3, as illustrated in Fig. 2(a). Due to the random and disordered friction between multiple rollers, there will be disharmony in the triboelectric layer regions corresponding to grid electrodes in different rollers. This disordered friction can result in output performance canceling each other, leading to a limit in electrical output after multiple work units are connected in parallel. Moreover, as the size of the work unit increases, the probability of this output performance canceling out each other increases, which seriously hinders the improvement of MR-TENG output performance. Therefore, it is necessary to improve the electrical output of MR-TENG by optimizing the energy management circuit system. The specific solution is to isolate interference between different work units through a rectification system of a single work unit. In a nutshell, we define parallel connection followed by rectification as PCR output mode, as exhibited in Fig. 2(b). Meanwhile, we define rectification followed by parallel connection as RPC output mode (Fig. 2(a)). Thereunder, we developed the differences in MR-TENG with different numbers of rollers' output performance between the two output modes. The motion frequency and maximum motion distance are 4 Hz and 5 cm, respectively. According to results in Fig. 2(c), under PCR output mode, the $V_{oc}$ of MR-TENG with one working (unit No. 1) can reach ~168 V, the $V_{oc}$ of MR-TENG with two working units (No. 1 and No. 2) can reach ~80 V, and the $V_{oc}$ of MR-TENG with working unit (No. 1, No. 2, and No. 3) can reach ~59.7 V, respectively. As the number of rollers increases, the $V_{oc}$ of MR-TENG shows a sharp downward trend, which is caused by the asynchronous operation between each working unit. According to results in Fig. 2(d), the same performance decline trend is reflected in the $I_{sc}$ of MR-TENG, and $I_{sc}$ decreases from ~2.06 μA to ~1.3 μA. From the results in Fig. 2(f), under RPC output mode, the $V_{oc}$ of MR-TENG with one working (unit No. 1) can reach ~143 V, the $V_{oc}$ of MR-TENG with two working units (No. 1 and No. 2) can reach ~147 V, and the $V_{oc}$ of MR-TENG with working unit (No. 1, No. 2, and No. 3) can reach ~166 V, respectively. Obviously, in RPC output mode, the $V_{oc}$ of MR-TENG demonstrates a trend of performance enhancement with the increase of working units. And this is because first rectification can overcome the problem of output asynchrony between different working units, resulting in enhanced output performance. According to results in Fig. 2(g), the same growth trend also occurred in the $I_{sc}$ of MR-TENG, and $I_{sc}$ grows from ~1.87 μA to ~2.17 μA. To further evaluate the energy differences generated by the two output modes, we charged the capacitor of 1 μF with MR-TENG under the same conditions, and indirectly calculated the generated energy by measuring the voltage value within the same time period. From the results in Fig. 2(e), under PCR output mode, the 1 μF capacitor can be charged to 20.7 V within 100 s, while under RPC output mode, the 1 μF capacitor can be charged to 34.7 V within 100 s. Through calculation in Fig. 2(h), under RPC output mode, MR-TENG can generate 602.045 μJ of electrical energy in 100 s, while under PCR output mode, MR-TENG can generate 214.245 μJ of electrical energy. Compared to the PCR output mode, the RPC output mode can achieve 2.8 times the electrical energy output, which indicates that the RPC output mode has higher superiority. By addressing the issue of mutual interference between different work units, the total energy output of MR-TENG has been significantly improved. Specifically, as the spatial utilization of TENG devices increases, the coupling problem of multiple working units is widespread. This solution approach has the opportunity to be applicable to solving the mutual interference problem of multiple working units in other TENG design processes, and has universal characteristics.

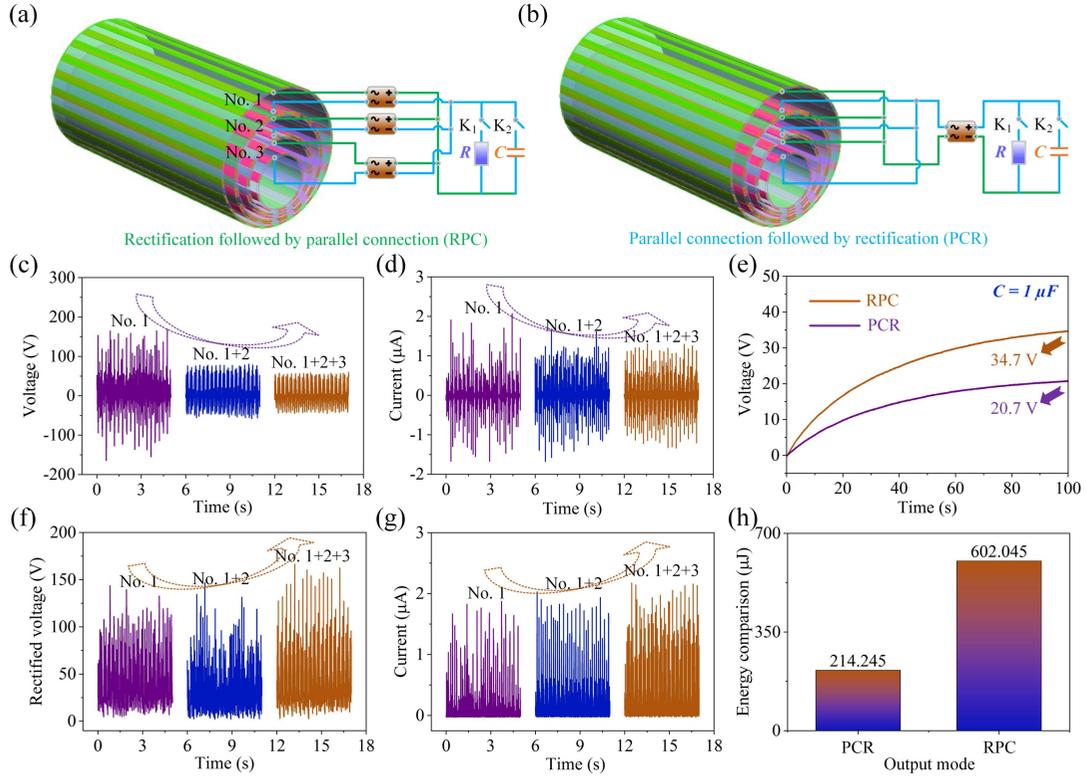

Fig. 2. The (a) RPC output mode and (b) PCR output mode of MR-TENG. The (c) $V_{oc}$ and (d) $I_{sc}$ of MR-TENG with different number of working unit. (e) The charging curves of MR-TENG for 1 μF capacitor through two output modes. The (f) $V_{oc}$ and (g) $I_{sc}$ of MR-TENG with different number of working unit under RPC output mode. (h) The electric energy stored in 1 μF capacitor generated by MR-TENG through two output mode.

### 3.3. The output performance of MR-TENG device.

Fig. 3(a, b) shows the $I_{sc}$ and $V_{oc}$ signal of MR-TENG under RPC output mode. According to the signal characteristics, the electrical output of the frequency increase effect can be observed, which will contribute to energy enhancement. The grid electrode brings an increase in frequency effect, which increases the number of contact separations between the two triboelectric layers, leading to performance growth. Furthermore, by studying the output current and voltage of MR-TENG under different resistance loads, the output power is calculated. From the results in Fig. 3(c), the maximum instantaneous power of MR-TENG can arrive at 405 μW (Matched load: 10 MΩ). As exhibited in Fig. 3(d), further reliability experiments have shown that MR-TENG can still maintain stable performance output under the condition of continuous operation of 6400 times (frequency: 4 Hz, time: 1600 s). According to SEM images of paper before and after the experiment in **Fig. S1 of Supplementary Materials**, as the triboelectric component of MR-TENG, the surface texture of the paper remains stable before and after 6400 cycles of MR-TENG operation, without any instability. Moreover, we investigated the energy output of each working unit of MR-TENG through relevant circuit systems in Fig. 3(e). Specifically, when switch $K_1$ is closed, working unit No. 1 charges the capacitor (4.7 µF), resulting in a voltage of 15.14 V within 100 s. When switching $K_1$ and $K_2$ are closed, working units No. 1 and No. 2 charge the capacitor (4.7 µF), leading to a voltage of 18.17 V within 100 s. When switch $K_1$, $K_2$, and $K_3$ are closed, working unit No. 1, No. 2, and No. 3 charge the capacitor (4.7 µF), leading to a voltage of 21.17 V within 100 s. Thus, as the number of work units increases, the utilization rate of MR-TENG space increases, while achieving an increase in output energy, indicating that this structural design and mode optimization can help enhance the roller structure TENG.

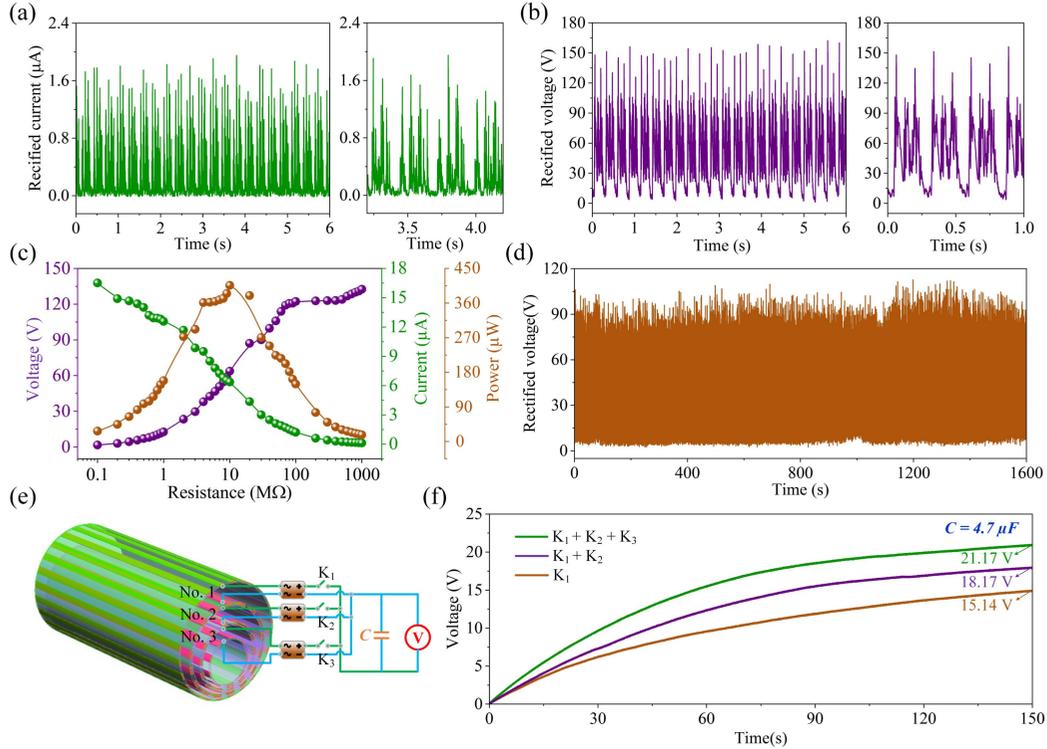

Fig. 3. The (a) $I_{sc}$ and (b) $V_{oc}$ of MR-TENG under RPC output mode. (c) The electrical output of MR-TENG. (d) Rectified voltage reliability test of MR-TENG. (e) The equivalent-circuit of MR-TENG under RPC output mode for output energy evaluation. (f) The charging curves of 4.7 µF capacitor charged by MR-TENG with different working units.

The contact frequency and maximum separation gap between the triboelectric layers in TENG devices will affect the output performance. When MR-TENG is subjected to mechanical impacts of different amplitudes and frequencies, the separation gap and friction frequency between the internal triboelectric layers of TENG devices will change accordingly. As illustrated in Fig. 4(a), in the case of frequency changes (from 1 Hz to 2 Hz), the rectified voltage peak value of MR-TENG under RPC output mode represents a slight increase (from ~126 V to ~162 V). At high frequencies, the rolling amplitude of the roller in the MR-TENG increases due to inertia, but limited by the inner diameter of the largest external roller, the rectified voltage of MR-TENG can increase slightly. For the rectified current in Fig. 4(b), it reveals a growth trend (from ~0.46 µA to ~2.07 µA) with increasing working frequency. High frequencies can lead to friction between rollers at a greater distance, resulting in high-frequency currents, and the rectified output current peak value will also increase accordingly. High mechanical frequency shocks will enhance the efficiency of triboelectric charge transfer, while high-density grid electrodes will further increase the frequency of output current, which will help enhance power generation efficiency. Moreover, as the amplitude increases, the output performance of MR-TENG also increases. As exhibited in Fig. 4(c), when the mechanical system provides motion with an amplitude of 3 cm to 10 cm, the rectified voltage of MR-TENG can rise from ~91 V to ~133 V under working frequency of 2 Hz, which is due to the large rolling amplitude of the roller caused by high amplitude. However, due to the limited spacing between the triboelectric layers inside MR-TENG, as the mechanical amplitude increases, the rectified $V_{oc}$ of MR-TENG will first increase and then tend to be relatively stable. As shown in Fig. 4(c), the rectified current of MR-TENG can grow from ~0.34 μA to ~0.87 μA, when the amplitude increases. The increase in rectified current of MR-TENG originates from the efficient relative motion between triboelectric layers. Specifically, a higher vibration amplitude will intensify the relative motion between the rollers, thereby improving the friction efficiency between the triboelectric layers and leading to an increase in rectified $I_{sc}$ of MR-TENG.

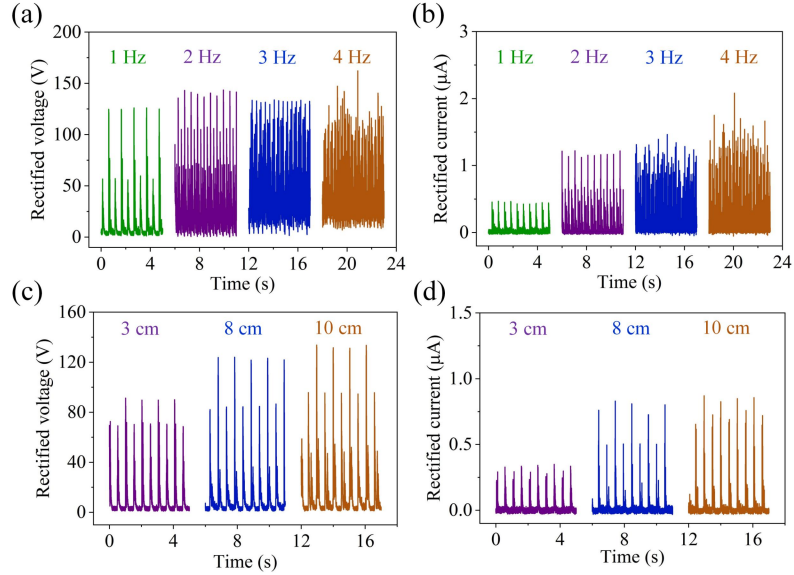

Fig. 4. The (a) rectified voltage and (b) rectified current of MR-TENG under various working frequencies. The (c) rectified voltage and (d) rectified current of MR-TENG under various movement distances.

The electrical energy generated by a single MR-TENG device is often difficult to meet the demand, so it is necessary to design an array that integrates multiple MR-TENG devices. Obviously, large-scale MR-TENG integration also faces energy management issues, mainly focusing on the connection methods between various MR-TENGs, such as parallel and series. Obviously, when connecting multiple TENG networks, the selection of connection methods is crucial. Accordingly, we investigated the output characteristics of the MR-TENG array composed of three MR-TENGs in parallel and series, respectively. In parallel mode, as the number of parallel MR-TENGs increases, the highest value of $V_{oc}$ shows a gentle trend (from ~164 V to ~166 V), while the lowest value of $V_{oc}$ tends to rise (from ~9.2 V to ~38.8 V), as illustrated in Fig. 5(a). Hence, the amplitude of the change in $V_{oc}$ decreases, which helps to provide a stable voltage for electrical appliances. For $I_{sc}$ of MR-TENG array in Fig. 5(b), the highest value of $I_{sc}$ exhibits a significant increase trend (from ~2 μA to ~5.39 μA), and the growth rate is 169.5%, which can be attributed to the increase in the total amount of transferred charges after parallel connection of MR-TENG, which in turn forces the $I_{sc}$ to increase. As shown in Fig. 5(c), further charging experiments showed that as the number of MR-TENGs in parallel increased, the voltage on 1 μF capacitor reached 53 V, 66 V, and 75 V within 100 s, respectively. Thus, the parallel connection method can improve the charging efficiency of the MR-TENG array. And through parallel integration, the output voltage stability of MR-TENG integration can be improved, which is very necessary for electrical appliances that need stable power supply. At the same time, it has also achieved a significant increase in output current. Moreover, for series mode, when the number of MR-TENG increases, unlike parallel mode, the highest value of $V_{oc}$ illustrates an upward trend (from ~168 V to ~197 V), while the lowest value of $V_{oc}$ has a smaller increase (from ~9.3 V to ~15. 8 V), as exhibited in Fig. 5(d). For Isc of the MR-TENG array in Fig. 5(e), as the number of MR-TENG increases, the highest value of $I_{sc}$ exhibits a significant increase trend (from ~2 μA to ~2.5 μA), and the growth rate is 25% lower than 169.5% in parallel mode. As shown in Fig. 5(f), as the number of MR-TENGs in series increased, the voltage on 1 μF capacitor charged by MR-TENG array can arrive at 53 V, 61 V, and 69 V within 100 s, respectively. Compared to the parallel mode, the charging effect of the MR-TENG array on the capacitor is reduced, which is due to the decrease in charge output caused by the series mode. Meanwhile, in series mode, a larger number of MR-TENG can achieve a higher stable voltage for the capacitor, which is due to the increase in the $V_{oc}$ of MR-TENG in series mode. In some special environments, high output voltage is required, and the $V_{oc}$ of multiple MR-TENGs can be strengthened through series connection to meet electricity demand. Thus, it is necessary to confirm the output mode of the MR-TENG array in specific application scenarios to achieve the maximum output function of the MR-TENG array. Fig. 5(g) illustrates the demonstration scenario of the MR-TENG array and linear dynamic system. The MR-TENG array can power 40 blue commercial light-emitting diodes (LEDs) under micro-vibration (mechanical frequency: 4 Hz, movement distance: 5 cm), as shown in Fig. 5(h).

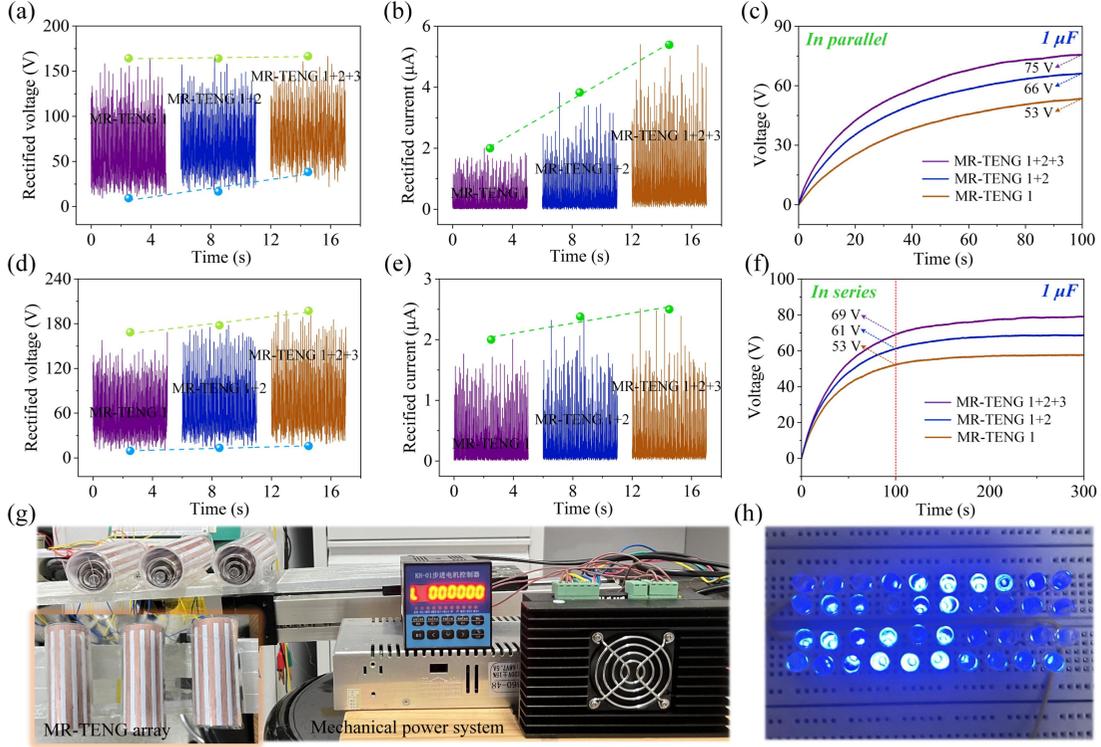

Fig. 5. The (a) $V_{oc}$ and (b) $I_{sc}$ of MR-TENG array composed of various MR-TENGs in parallel. (c) The charging curves of 1 μF commercial capacitor charged by MR-TENG array composed of various MR-TENGs in parallel. The (d) $V_{oc}$ and (e) $I_{sc}$ of MR-TENG array composed of various MR-TENGs in series. (f) The charging curves of 1 μF commercial capacitor charged by MR-TENG array composed of various MR-TENGs in series. (g) The demonstration scenario of MR-TENG array and linear dynamic system. (h) 40 commercial LEDs powered by the MR-TENG array.

## 4. Conclusion

In summary, we propose a MR-TENG based on multi-roller structure for water wave energy harvesting. Compared to the previous roller structure TENG, the MR-TENG achieved higher space utilization and bring higher mechanical energy harvesting efficiency. Additionally, the connection method between MR-TENG work units will determine the output energy of MR-TENG, such as rectification followed by parallel connection (RPC) mode and parallel connection followed by rectification (PCR) mode. According to charging results for 1 μF using MR-TENG, MR-TENG can generate 602.045 μJ energy in RPC mode, which is 2.8 times higher than 214.245 μJ in PCR mode, which is attributed to the asynchronous work rhythm between each work unit in MR-TENG. The maximum instantaneous power of MR-TENG can arrive at 405 μW (Matched load: 10 MΩ). The $I_{sc}$ and $V_{oc}$ of MR-TENG can arrive at ~2.06 μA and ~166 V, respectively. We also investigated the influence of connection mode of MR-TENG array on electrical output, such as parallel and series. In parallel mode, the output energy efficiency of the MR-TENG array is relatively high. In series mode, the $V_{oc}$ of the MR-TENG array is higher. Moreover, the MR-TENG array can power 40 blue commercial light-emitting diodes (LEDs) under micro-vibration. At present, further exploration is still needed for MR-TENG devices, especially in marine environments, to improve their output performance and optimize their structure and materials.

## 5. Acknowledgments

The authors would like to thank the Ocean Science Experimental Teaching Center of Zhejiang University for SEM characterization.

## References

[1] Liguo X, Ahmad M, Khattak S I. Impact of innovation in marine energy generation, distribution, or transmission-related technologies on carbon dioxide emissions in the United States. Renewable and Sustainable Energy Reviews, 2022, 159: 112225.
[2] Chang A, Uy C, Xiao X, et al. Self-powered environmental monitoring via a triboelectric nanogenerator. Nano Energy, 2022, 98: 107282.
[3] Chao S, Ouyang H, Jiang D, et al. Triboelectric nanogenerator based on degradable materials. EcoMat, 2021, 3(1): e12072.
[4] Cheng T, Shao J, Wang Z L. Triboelectric nanogenerators. Nature Reviews Methods Primers, 2023, 3(1): 39.
[5] Rodrigues C, Nunes D, Clemente D, et al. Emerging triboelectric nanogenerators for ocean wave energy harvesting: state of the art and future perspectives. Energy & Environmental Science, 2020, 13(9): 2657-2683.
[6] Zhang C, Liu Y, Zhang B, et al. Harvesting wind energy by a triboelectric nanogenerator for an intelligent high-speed train system. ACS Energy Letters, 2021, 6(4): 1490-1499.
[7] He L, Zhang C, Zhang B, et al. A dual-mode triboelectric nanogenerator for wind energy harvesting and self-powered wind speed monitoring. ACS nano, 2022, 16(4): 6244-6254.
[8] Zhao B, Li Z, Liao X, et al. A heaving point absorber-based ocean wave energy convertor hybridizing a multilayered soft-brush cylindrical triboelectric generator and an electromagnetic generator. Nano Energy, 2021, 89: 106381.
[9] Zhao T, Xu M, Xiao X, et al. Recent progress in blue energy harvesting for powering distributed sensors in ocean. Nano Energy, 2021, 88: 106199.
[10] Chen F, Zhang S, Guan P, et al. High-Performance Flexible Graphene Oxide-Based Moisture-Enabled Nanogenerator via Multilayer Heterojunction Engineering and Power Management System. Small, 2023: 2304572.
[11] Deng W, Zhou Y, Libanori A, et al. Piezoelectric nanogenerators for personalized healthcare. Chemical Society Reviews, 2022, 51(9): 3380-3435.
[12] Wang Z L. Triboelectric nanogenerator (TENG)—sparking an energy and sensor revolution. Advanced Energy Materials, 2020, 10(17): 2000137.
[13] Zhang D, Wang Y, Yang Y. Design, performance, and application of thermoelectric nanogenerators. Small, 2019, 15(32): 1805241.

[14] Wang H, Xu L, Bai Y, et al. Pumping up the charge density of a triboelectric nanogenerator by charge-shuttling. Nature Communications, 2020, 11(1): 4203.

[15] He W, Liu W, Chen J, et al. Boosting output performance of sliding mode triboelectric nanogenerator by charge space-accumulation effect. Nature communications, 2020, 11(1): 4277.

[16] Zhou Q, Pan J, Deng S, et al. Triboelectric nanogenerator-based sensor systems for chemical or biological detection. Advanced Materials, 2021, 33(35): 2008276.

[17] Zhang J, Sun Y, Yang J, et al. Irregular wind energy harvesting by a turbine vent triboelectric nanogenerator and its application in a self-powered on-site industrial monitoring system. ACS applied materials & interfaces, 2021, 13(46): 55136-55144.

[18] Xu S, Liu G, Wang J, et al. Interaction between water wave and geometrical structures of floating triboelectric nanogenerators. Advanced Energy Materials, 2022, 12(3): 2103408.

[19] Wang Z L. On the expanded Maxwell's equations for moving charged media system–General theory, mathematical solutions and applications in TENG. Materials Today, 2022, 52: 348-363.

[20] Cui S, Zhou L, Liu D, et al. Improving performance of triboelectric nanogenerators by dielectric enhancement effect. Matter, 2022, 5(1): 180-193.

[21] Gao Q, Cheng T, Wang Z L. Triboelectric mechanical sensors—Progress and prospects. Extreme Mechanics Letters, 2021, 42: 101100.

[22] Xia K, Fu J, Xu Z. Multiple-frequency high-output triboelectric nanogenerator based on a water balloon for all-weather water wave energy harvesting. Advanced Energy Materials, 2020, 10(28): 2000426.

[23] Xia K, Wu D, Fu J, et al. A high-output triboelectric nanogenerator based on nickel-copper bimetallic hydroxide nanowrinkles for self-powered wearable electronics. Journal of Materials Chemistry A, 2020, 8(48): 25995-26003.

[24] Xia K, Xu Z, Hong Y, et al. A free-floating structure triboelectric nanogenerator based on natural wool ball for offshore wind turbine environmental monitoring. Materials Today Sustainability, 2023: 100467.

[25] Xu L, Wu H, Yao G, et al. Giant voltage enhancement via triboelectric charge supplement channel for self-powered electroadhesion. ACS nano, 2018, 12(10): 10262-10271.

[26] Cheng P, Guo H, Wen Z, et al. Largely enhanced triboelectric nanogenerator for efficient harvesting of water wave energy by soft contacted structure. Nano Energy, 2019, 57: 432-439.

[27] Liang X, Jiang T, Liu G, et al. Triboelectric nanogenerator networks integrated with power management module for water wave energy harvesting. Advanced Functional Materials, 2019, 29(41): 1807241.

[28] Lei R, Zhai H, Nie J, et al. Butterfly-Inspired Triboelectric Nanogenerators with Spring-Assisted Linkage Structure for Water Wave Energy Harvesting. Advanced Materials Technologies, 2019, 4(3): 1800514.

[29] Wu Z, Guo H, Ding W, et al. A hybridized triboelectric–electromagnetic water wave energy harvester based on a magnetic sphere. ACS nano, 2019, 13(2): 2349-2356.

[30] Zhang X, Yu M, Ma Z, et al. Self-powered distributed water level sensors based on liquid–solid triboelectric nanogenerators for ship draft detecting. Advanced Functional Materials, 2019, 29(41): 1900327.

[31] Feng Y, Liang X, An J, et al. Soft-contact cylindrical triboelectric-electromagnetic hybrid

nanogenerator based on swing structure for ultra-low frequency water wave energy harvesting. Nano Energy, 2021, 81: 105625.

[32] Cao B, Wang P, Rui P, et al. Broadband and Output-Controllable Triboelectric Nanogenerator Enabled by Coupling Swing-Rotation Switching Mechanism with Potential Energy Storage/Release Strategy for Low-Frequency Mechanical Energy Harvesting. Advanced Energy Materials, 2022, 12(46): 2202627.

[33] Han J, Liu Y, Feng Y, et al. Achieving a Large Driving Force on Triboelectric Nanogenerator by Wave-Driven Linkage Mechanism for Harvesting Blue Energy toward Marine Environment Monitoring. Advanced Energy Materials, 2023, 13(5): 2203219.

[34] Hu Y, Qiu H, Sun Q, et al. Wheel-structured Triboelectric Nanogenerators with Hyperelastic Networking for High-Performance Wave Energy Harvesting. Small Methods, 2023: 2300582.